\documentclass{jfm}

\usepackage{graphicx}
\usepackage{newtxtext}
\usepackage{newtxmath}
\usepackage{natbib}
\usepackage{hyperref}
\hypersetup{
    colorlinks = true,
    urlcolor   = blue,
    citecolor  = black,
}

\title{A predictive model of drop breakup in turbulence}

\author{Daniel Mor\'{o}n\aff{1,2}
  \corresp{\email{dmoron@ing.uc3m.es}},
  Ianto Cannon\aff{1},
  Alberto Vela-Mart\'{i}n\aff{2}
 \and Marc Avila\aff{1,3}}

\affiliation{
\aff{1}University of Bremen, Center of Applied Space Technology and Microgravity (ZARM), Am Fallturm 2, 28359 Bremen, Germany.
\aff{2}Universidad Carlos III de Madrid, Escuela Politécnica Superior, Department of Aerospace Engineering,  28911 Leganés, Spain.
\aff{3}University of Bremen, MAPEX Center for Materials and Processes, Am Biologischen Garten 2, 28359 Bremen, Germany.}

\makeatletter
\def\ps@titlepage{\leftskip\z@\let\@mkboth\@gobbletwo\vfuzz=5\p@
  \def\@oddhead{\@j@urnal\hfil\llap{\thepage}}%
  \def\@evenhead{\@j@urnal\hfil\llap{\thepage}}%
  \def\@oddfoot{}\def\@evenfoot{}%
  \def\sectionmark##1{}\def\subsectionmark##1{}}
\def\pagelimitfooter{}
\def\absfooterflag{}
\makeatother

\begin{document}
\maketitle

%%%%%%%%%%%%%%%%%%%%%%%%%%%%%%%%%%%%%%%%%%%%%%%%%%%%%%%%%%%%%%%%%%%%%%%%%
% ABSTRACT
%%%%%%%%%%%%%%%%%%%%%%%%%%%%%%%%%%%%%%%%%%%%%%%%%%%%%%%%%%%%%%%%%%%%%%%%%
\begin{abstract}
Reduced-order models of drop breakup in turbulence are highly sought after, but the bidirectional, nonlinear coupling between the interfacial and hydrodynamic stresses is an obstacle to their development. By leveraging a decomposition of the flow into outer and inner regions introduced by \citet{AlbertoJFM}, we show that drop deformation is driven by outer eddies that stretch the fluid-fluid interface due to the non-local 
coupling between strain and vorticity. Capillary forces oppose stretching and, as the drop relaxes, surface energy is transferred back to the velocity field through the genesis of inner eddies (inside the drop and its boundary layer). Hence, for breakup to occur, stretching events caused by outer eddies must inject energy faster than the interface can convert it into inner eddies. Direct numerical simulations initialised with ellipsoidal drops reveal that the energy transfer to the inner eddies is governed by the capillary time, whereas the outer forcing acts on the eddy-turnover time scale. Building on these observations and starting from the governing equations, we derive a simple model for the breakup rate. Although the underlying assumptions are strictly valid only in the limit of large surface tension, the resulting equation quantitatively predicts drop breakup rates over a wide range of Weber numbers. All model parameters can be measured a priori and quantify properties of the outer turbulent flow and the drop response. This work establishes a link between extreme outer-stretching events in turbulence and the memoryless nature of drop breakup at low Weber numbers.
\end{abstract}


%%%%%%%%%%%%%%%%%%%%%%%%%%%%%%%%%%%%%%%%%%%%%%%%%%%%%%%%%%%%%%%%%%%%%%%%%
% INTRODUCTION
%%%%%%%%%%%%%%%%%%%%%%%%%%%%%%%%%%%%%%%%%%%%%%%%%%%%%%%%%%%%%%%%%%%%%%%%%
\section{Introduction}
% General description of the problem 
The size of drops and bubbles regulates the efficiency of liquid-gas exchange at the ocean-atmosphere interface \citep{deane2002scale,deike2022mass}, biodegradation in oil spills \citep{brizzolara2024immiscible}, emulsification of foods and chemicals \citep{mathijssen2023culinary,haakansson2019emulsion}, and atomisation of liquids \citep{jiang2010physical}. The particle size distribution controls the rates of heat, mass, and momentum transfer, as well as chemical reaction rates. The key mechanisms driving the particle size distribution are breakup and coalescence. In turbulent flows, breakup is dominated by inertial-range eddies, with coalescence becoming relevant only at low turbulent intensities, even at high concentrations \citep{hinze1955fundamentals,Mukherjee2019,yi2021global}. The most important parameter describing the breakup process is the breakup rate or frequency, $\kappa$, yet there is no clear method to determine it experimentally. Even in single-particle experiments \citep{eastwood2004breakup,andersson2006breakup,maass2012,solsvik2015single,beckedorff2025jet}, the difficulties arise from limited observation windows and statistics, and inevitable flow inhomogeneities and anisotropies~\citep{haakansson2020validity}. As a result, measured breakup rates and model predictions can differ by orders of magnitude \citep{qi2020towards,ni2024deformation}.

% Phenomenological approach
The general phenomenological approach to turbulent breakup was introduced independently by \citet{kolmogorov1949disintegration} and \citet{hinze1955fundamentals}. They considered the exchange of kinetic and surface energies, which, in the inertial range of turbulence, is governed by the Weber number,
\begin{align}
    \mathrm{We}=\frac{\rho \, \varepsilon^{2/3} \, d^{5/3}}{\sigma}\text{.}
\end{align}
Here, $\rho$, $\varepsilon$, $d$ and $\sigma$ are the density of the carrier fluid, the rate of energy dissipation per unit mass, the particle diameter and  the surface tension, respectively. \citet{kolmogorov1949disintegration} and \citet{hinze1955fundamentals} assumed breakup to occur when the kinetic energy of velocity fluctuations of size $d$ exceeds the surface energy, i.e. $\mathrm{We} > 1$. \citet{coulaloglou1977description} modelled breakup as a Poisson process with a breakup rate proportional to the probability that the drop encounters a velocity fluctuation of sufficiently high kinetic energy. By assuming an exponential distribution of turbulent energy fluctuations, they arrived at
\begin{align}
    \kappa \approx c_{1}\exp\left(-2~c_{2}~\mathrm{We}^{-1}\right) \text{,} \label{eq:C77}
\end{align}
where $c_{1}$ and $c_{2}$ are model constants. Their model has been widely used and extended in the literature; however, the measured values of $c_1$ and $c_2$ vary greatly across studies \citep{vankova2007,maass2012,solsvik2015single}. Recently, \citet{AlbertoSciAdv} performed thousands of independent, fully resolved simulations of drop breakup in the idealised case of homogeneous isotropic turbulence (HIT) and showed that drop breakup is indeed a memoryless (Poisson) process. Their computed breakup rates were in excellent agreement with equation~\eqref{eq:C77}, with the parameters $c_1=14.8$ and $c_2=3.9$ obtained from a least-squares fit to their data, which covered $\mathrm{We}\in[1.45,7.27]$, Reynolds number based on the Taylor microscale $\mathrm{Re}_{\lambda}=\{31,58,96\}$ and viscosity ratios $0.5$, $1$ and $2$.

The model proposed by \citet{coulaloglou1977description} implicitly assumes that the energy of the eddies encountered by the drop is fully converted into surface energy before the drop can relax, generating in turn eddies that dissipate part of the energy input. Recently, \citet{qi2024breaking} showed that breakup can only occur when the energy injection rate matches or exceeds the dissipation rate. For small drop oscillations in a quiescent fluid \citep{lamb1932hydrodynamics}, dissipation occurs in the particle boundary layer due to viscosity \citep{reid1960oscillations,miller1968oscillations}. Turbulence enhances this dissipation, resulting in an effective viscosity 5--8 times larger than the corresponding laminar value \citep{xiao2021impact}. This effect is commonly modelled as an eddy viscosity based on the velocity and length of inertial eddies of size $d$ \citep{riviere2024bubble,qi2024breaking}. 

In this paper, we derive a simple model of drop breakup in the spirit of \citet{coulaloglou1977description}, by formalising their phenomenological approach and accounting for the energy extracted from the interface during drop relaxation. The paper is structured as follows. In \S\ref{sec:Methods}, we briefly describe the numerical methods, introduce the energy balance and review the decomposition of \citet{AlbertoJFM} to separate the contributions of inner/outer eddies (vorticity) to the interfacial strain in the equation for the surface energy. In \S\ref{sec:strain}, we exploit their decomposition to investigate the temporal dynamics of the breakup process. Specifically, we show that at low Weber numbers inner eddies are generated in response to the stretching caused by outer eddies. Additionally, the transfer from surface energy into inner eddies is shown to occur at the capillary time scale. We demonstrate that this energy transfer can be modelled in the equation for the surface energy as a linear damping term with a constant damping rate. In \S\ref{sec:model}, we use these results to derive a stochastic differential equation for the surface energy, in which a forcing term solely depending on the turbulence characteristics (the stretching by outer eddies) causes deformation at the eddy-turnover time scale, and a damping term (the energy extraction through the genesis of inner eddies) relaxes the drop toward equilibrium at the capillary time scale.  By taking the limit of breakup caused by single, extreme events of outer stretching, we obtain an analytical equation that matches the simplicity of equation~\eqref{eq:C77} and admits an a priori estimation of the model constants. In \S\ref{sec:Conclusion}, we critically assess the model and discuss its validity beyond the flow conditions studied here.

%%%%%%%%%%%%%%%%%%%%%%%%%%%%%%%%%%%%%%%%%%%%%%%%%%%%%%%%%%%%%%%%%%%%%%%%%
% METHODS
%%%%%%%%%%%%%%%%%%%%%%%%%%%%%%%%%%%%%%%%%%%%%%%%%%%%%%%%%%%%%%%%%%%%%%%%%
\section{Methods}\label{sec:Methods}

\subsection{Direct numerical simulations of drop breakup}

Throughout the paper, we consider two immiscible incompressible fluids with equal viscosity and density in a triply periodic cubic box of length $L$. The two-phase flow is simulated by solving the coupled Navier--Stokes and Cahn--Hilliard equations \citep{jacqmin1999calculation}. In this approach, an infinitely differentiable phase-field variable distinguishes between the two immiscible fluids and changes smoothly across a finite-thickness interface. The evolution of the phase-field variable is governed by the Cahn--Hilliard equation, which features two dimensionless parameters. First, the Cahn number, $\mathrm{Cn}$, which is the dimensionless interface thickness (scaled with the integral length of the flow, $L$). Second, the P\'eclet number, $\mathrm{Pe}$, which controls the interface mobility. \citet{magaletti2013sharp} performed an asymptotic analysis of the Navier--Stokes and Cahn--Hilliard equations and showed that if $\mathrm{Pe}$ is chosen inversely proportional to $\mathrm{Cn}$, the interface and bulk dynamics are matched and the correct surface tension is recovered in the Navier--Stokes equations. The sharp-interface limit is rigorously recovered asymptotically as $\mathrm{Cn}\rightarrow0$. An advantage of this formulation is that topological changes of the flow (e.g. in drop breakup) are handled consistently as $\mathrm{Cn}$ is reduced. 

The flow is sustained by a constant-power forcing at the largest scales ($0<|\pmb{k}|<2$) \citep{cardesa}, so that in the statistically steady state the time-averaged dissipation rate $\varepsilon$ equals the prescribed energy injection rate. The equations are discretised using a pseudo-spectral method with a Fourier basis comprising $N/2$ modes in each spatial direction, where $N=256$ denotes the number of grid points in physical space. Nonlinear terms are fully de-aliased using the $2/3$ truncation rule, and temporal advancement is performed with a third-order low-storage Runge--Kutta scheme. Additional details regarding the numerical methods and their validation can be found in \citet{AlbertoJFM}.

\begin{table}
\centering
\caption{Direct numerical simulations carried out in this paper.}
\label{tab:Database}
\begin{tabular}{c|ccccccc}
\hline
Weber number & 1.2 & 1.3 & 1.4 & 1.6 & 1.8 & 2.9 & 3.6 \\ 
\hline
Number of DNS & 180 & 200 & 200 & 200 & 200 & 500 & 600 \\
\hline
\end{tabular}
\end{table}

\begin{figure}
    \centering
    \includegraphics[width=\textwidth]{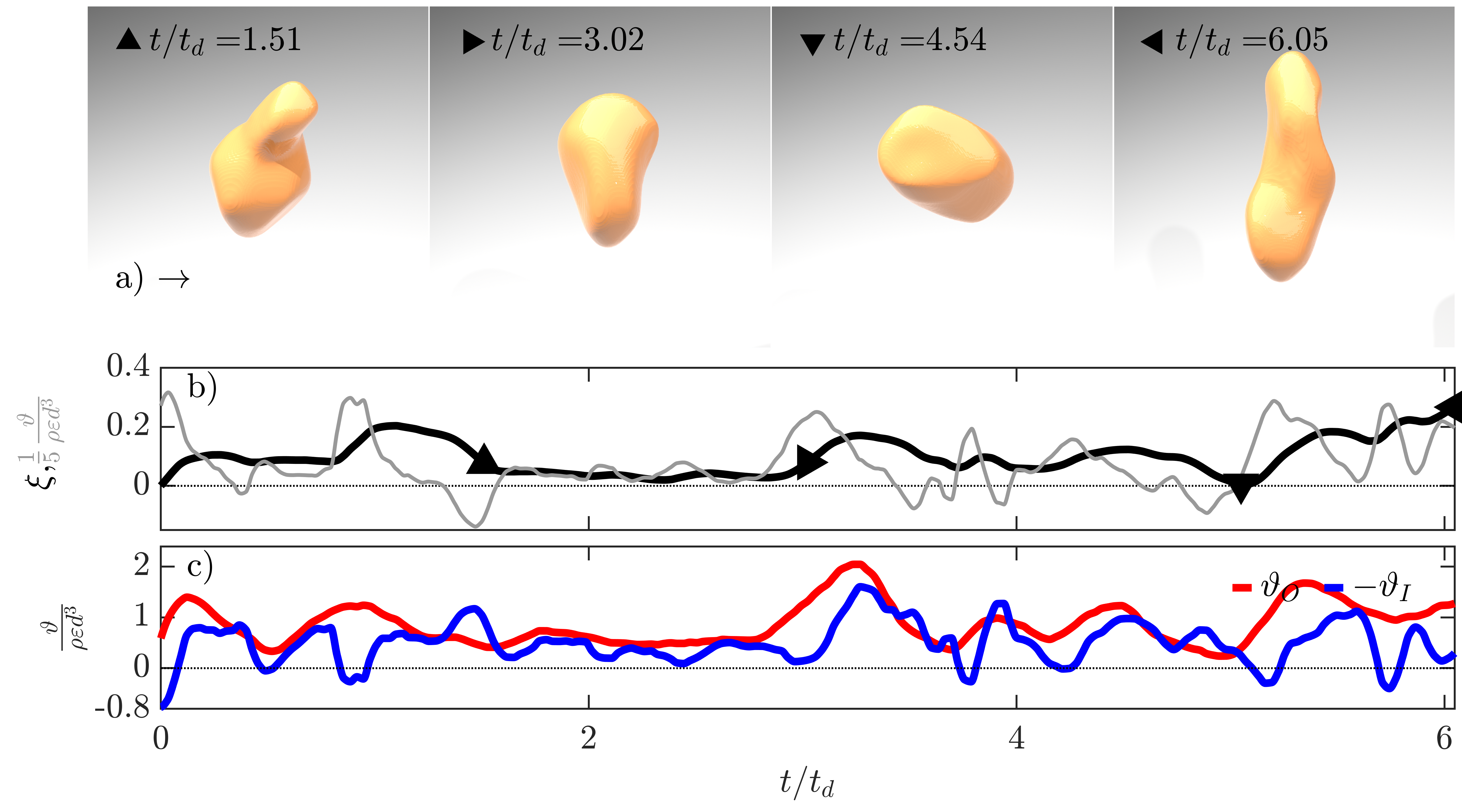}
    \caption{a) Snapshots of a drop deforming in a turbulent field at $\mathrm{We}=1.8$. Times are normalised with the eddy-turnover time at the drop scale, $t_{d}=(d^{2}/\varepsilon)^{1/3}$. b) Temporal evolution of the normalised excess surface energy $\xi$, equation~\eqref{eq:xi}, and the strain $\vartheta$, equation~\eqref{eq:stretch}, scaled by 1/5 for better visualisation. The black markers denote the times of the snapshots in a). c) Temporal evolution of the outer and inner strain. Note that the inner strain is plotted with an opposite sign.}
    \label{fig:fig1}
\end{figure}

In table~\ref{tab:Database}, we enumerate the direct numerical simulations carried out and analysed in this paper. Following \citet{AlbertoSciAdv}, each simulation was initialised by introducing a spherical drop of diameter $d=0.3L$ into a fully developed turbulent flow field. Each turbulent flow field is initialised with a random initial condition and the drop is introduced after the velocity field reaches a statistically steady state. The governing flow parameters are $\mathrm{Re}_{\lambda}\approx58$, $\mathrm{Cn}\lesssim0.009$ and $\mathrm{Pe}=1/(3~\mathrm{Cn})$; \citet{AlbertoSciAdv} showed that doubling the interface thickness, $\mathrm{Cn}$, and accordingly halving the numerical resolution in each direction does not affect breakup statistics. However, it increases mass leakage \citep{yue2007spontaneous} and practically limits the duration of the runs \citep{AlbertoSciAdv}. To alleviate this problem, we fixed $\mathrm{Cn}\lesssim0.009$. 

\subsection{Energy balance}
Snapshots of a drop interacting with turbulence are shown in figure~\ref{fig:fig1}a, and movies of drop deformation and breakup are provided in the supplementary material. We consider the global exchange of kinetic $E=\langle\frac{\rho}{2}\,u_i u_{i}\rangle_V$ and surface $\mathcal{H}=\sigma A$ energies, where $A$ is the drop's surface area. Volume-integration of the governing equations yields the following equations for the energy balance \citep{Pope1988,DoddFerrante2016,AlbertoJFM,VaheThiesset2026}:
\begin{align}
    \mathrm{d}_{t}E&=\mathcal{P} -\epsilon- \vartheta \text{,}
    \label{eq:EE} \\
    \mathrm{d}_{t} \mathcal{H}&=\vartheta  \text{,}
    \label{eq:OGH}
\end{align}
where $\mathcal{P}=\langle u_{i} f_{i} \rangle_{V}$ is the power input, $f_i$ the forcing that drives the turbulent flow, $\epsilon=2 \langle \mu  S_{ij}S_{ij}  \rangle_{V}=\rho V \varepsilon$ the dissipation rate and $\mu$ the dynamic viscosity. The energy exchange is controlled by the coupling (strain) term 
\begin{equation}\label{eq:stretch}
    \vartheta = - \left \langle \sigma n_{i} S_{ij} n_{j} \right \rangle_{S} \text{,}
\end{equation}
where $n_i$ is the normal vector to the drop surface and $\left \langle \bullet \right \rangle_{S}$ denotes integration over the interface between the two fluids. We stress that equations~\eqref{eq:EE}--\eqref{eq:stretch} are exact in the sharp-interface limit.

In figure~\ref{fig:fig1}b, we show the temporal evolution of the excess surface energy of the drop, 
\begin{align}\label{eq:xi}
    \xi=\frac{\mathcal{H}}{\mathcal{H}_{0}}-1 \text{,}
\end{align}
where $\mathcal{H}_0=\sigma\pi d^2$ is the surface energy of the spherical drop. For simulations with an initial spherical drop, $\mathcal{H}\left(0\right)=\mathcal{H}_{0}$. Times are scaled with the eddy-turnover time at the drop scale, $t_{d}=(d^{2}/\varepsilon)^{1/3}$. The dynamics of $\xi$ follow the dynamics of the strain. Specifically, positive values of $\vartheta$ (interface stretching) imply a net conversion rate of kinetic energy into surface energy and lead to an increasing $\mathcal{H}$, whereas negative values (interface compression) correspond to the generation of eddies as the drop relaxes towards a spherical shape, $\mathcal{H}\rightarrow \mathcal{H}_0$. These dynamics are observed in figure~\ref{fig:fig1}b, where phases of increasing interfacial strain, $\vartheta$, are followed by phases of increasing excess surface energy, $\xi$.

\subsection{Decomposition of the strain by inner and outer eddies}

\label{sec:outer_inner}
We decompose the vorticity field, $\pmb{\omega} = \pmb{\nabla} \times \pmb{u}$, into inner and outer fields to separate their contributions to the interfacial strain. This decomposition is motivated by the non-local nature of the coupling between vorticity and strain in turbulence \citep{ohkitani1995nonlocal,hamlington2008local}, which implies that eddies outside the boundary layer around the drop may produce significant strain at the interface. Indeed, \citet{AlbertoJFM} showed that a substantial part of the strain at the interface is generated by outer vorticity.  For completeness, we briefly review the rationale behind their decomposition and its mathematical definition.

We consider a distance $\Delta$ from the drop interface and define the kernel $G(\pmb{x};\Delta)$, which equals zero at points $\pmb{x}$ inside the drop and at points outside the drop that lie within a distance $\Delta$ of the interface; and equals one elsewhere. The outer vorticity vector is first computed as $\pmb{\omega}'^{O}= G \left( \pmb{x}; \Delta \right) \pmb{\omega}$. However, this field is not divergence-free. To correct it, it is projected onto its closest divergence-free field, 
\[
\pmb{\omega}^{O}=\pmb{\omega}'^{O}-\pmb{\nabla}\psi,
\] 
where $\nabla^2 \psi =\nabla \cdot \pmb{\omega}'^{O}$ is solved subject to periodic boundary conditions. The strain on the drop induced by the outer vorticity field can be computed from the Biot--Savart law \citep{ohkitani1995nonlocal}. Specifically, by taking the curl of the vorticity and using  that $\pmb{\nabla} \cdot \pmb{u}^{O}=0$,  the outer velocity field can be retrieved by solving 
\[
\nabla^{2} \pmb{u}^{O}=-\pmb{\nabla} \times \pmb{\omega}^{O}\text{.}
\]
The outer strain tensor is then
\[
S_{ij}^{O}=\frac{1}{2}\left(\partial_{j}u_{i}^{O} + \partial_{i} u_{j}^{O} \right),
\] 
and the inner strain tensor $S_{ij}^{I}=S_{ij}-S_{ij}^{O}$. We fix $\Delta =0.19~d$, which ensures that the inner and outer fields are decorrelated \citep{AlbertoJFM}. $\Delta$ can be interpreted as the thickness of the boundary layer around the drop and is of similar magnitude to that reported for bubbles ($0.1~d$) by \citet{riviere2024bubble}. Note that $\pmb{\nabla} \times \pmb{\omega}^{O} \equiv \pmb{\nabla} \times \pmb{\omega}'^{O}$ since $\pmb{\nabla}\times\pmb{\nabla}\psi=0$. Although the divergence-free correction may appear redundant, it is retained as it allows us to obtain a physically consistent vorticity field.

By applying this decomposition to the interfacial strain term, equation~\eqref{eq:stretch}, the equation for surface energy, equation~\eqref{eq:OGH}, can be written as,
\begin{equation}\label{eq:master}
\mathrm{d}_{t} \mathcal{H}=\vartheta_O + \vartheta_I,
\end{equation}
which forms the basis for our model. In figure~\ref{fig:fig1}c, we show an example of the temporal evolution of the outer strain, $\vartheta_O$, and minus the inner strain, $-\vartheta_I$. We observe that: (i) increases of $-\vartheta_I$ follow increases of $\vartheta_O$; (ii) $\vartheta_O>0$ throughout the run, whereas $ \vartheta_I$ changes sign in time, but is negative on average. In the next section, we analyse their  statistics and dynamics for all the cases in our database.

%%%%%%%%%%%%%%%%%%%%%%%%%%%%%%%%%%%%%%%%%%%%%%%%%%%%%%%%%%%%%%%%%%%%%%%%%
% RESULTS    
%%%%%%%%%%%%%%%%%%%%%%%%%%%%%%%%%%%%%%%%%%%%%%%%%%%%%%%%%%%%%%%%%%%%%%%%%
\section{Dynamics of the interfacial strain by inner and outer eddies}\label{sec:strain}

\begin{figure}
    \centering
    \includegraphics[width=\textwidth]{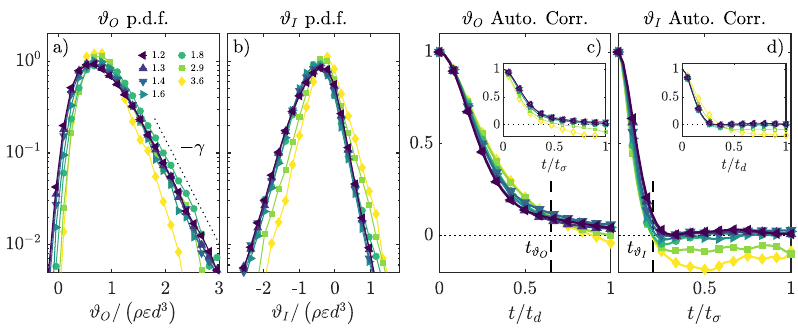}
    \caption{Statistics of the outer and inner strain. a) Probability density function of the outer strain at different Weber numbers (as indicated in the legend) normalised with inertial units. b) Same as a) but for the inner strain.  c) Temporal autocorrelation of the outer strain. d) Same as c) but for the inner strain, where $t_{\sigma}= \sqrt{\rho d^{3}/\sigma}$ is the capillary time. The insets in c and d show the autocorrelation function with respect to the alternative time scale $t_{\sigma}$ or $t_{d}$. With a vertical line we denote the corresponding autocorrelation time.}
    \label{fig:fig2}
\end{figure}

\citet{AlbertoJFM} showed that the outer strain is positive on average (interface stretching) and independent of the surface tension, whereas the inner strain is negative on average (interface compression) and depends both on the surface tension and the turbulence dynamics. Figures~\ref{fig:fig2}a and b extend their previous analysis to lower Weber numbers. At $\mathrm{We}<2$ the probability density functions of $\vartheta_{O}$ and $\vartheta_{I}$ are independent of the surface tension. The outer strain exhibits a pronounced positive exponential tail, with decay rate $\gamma\approx2.68$, which reflects the strong non-Gaussian behaviour and the prevalence of extreme events in the turbulent strain field \citep{jimenez1993structure}. The inner strain exhibits a pronounced negative tail, corresponding to strong relaxation events. As the Weber number increases, $\mathrm{We}\gtrsim2$, the positive tail of the inner strain grows and its mean becomes less negative. 

\subsection{Dynamics of the inner and outer strain}

We computed the autocorrelation functions of the outer, $\text{ACF}_{\vartheta_{O}} \left( t \right)$, and inner, $\text{ACF}_{\vartheta_{I}} \left( t \right)$, strain. To ensure a statistically steady state, we discarded the initial equilibration time of the simulations ($t \lesssim t_{d}$, see figure~\ref{fig:figHinit} in appendix~\ref{ap:Hinf}), and the final eddy-turnover time ($t \gtrsim t_{b}-t_{d}$) if the drop broke. Additionally, we did not assume the time signals to be stationary and normalised the autocorrelation functions so their maximum is $1$ at time $t = 0$. We computed the autocorrelation time as $t_{\vartheta_{O/I}}= \int_{0}^{\infty}\text{ACF}_{\vartheta_{O/I}} \left( t \right) \mathrm{d}t$.  In figure~\ref{fig:fig2}c, we show that the duration of typical events of outer strain is fairly constant among the Weber numbers considered here; it collapses better when scaled with $t_{d}$ than when scaled with the capillary time $t_{\sigma}=\sqrt{\rho d^{3}/\sigma}$. The corresponding autocorrelation time is $t_{\vartheta_{O}}\approx 0.65 t_{d}$. By contrast, the inner strain $\vartheta_I$ collapses slightly better with $t_{\sigma}$, with $t_{\vartheta_{I}} \approx 0.22~ t_{\sigma}$.

We observe that, as the Weber number increases, the inner strain autocorrelation does not converge to $0$ at large times, but to a negative finite value. At high Weber numbers, the capillary time scale becomes comparable to or larger than the mean breakup time ($t_{\sigma} \gtrsim 1.7t_d$ at $\mathrm{We}\approx 2.9$), resulting in increasingly short signals for the autocorrelation calculation. The negative autocorrelation values may be a consequence of these limited signal durations, which introduce a bias towards a few, unusually long-lived trajectories. Nevertheless, this behaviour may also have a physical origin: at high Weber numbers, the inner flow increasingly injects energy into the drop while remaining, on average, its only energy sink. Consequently, trajectories in which the inner flow more efficiently extracts surface energy may survive longer, potentially biasing the autocorrelation towards negative values.

\begin{figure}
    \centering
    \includegraphics[width=\textwidth]{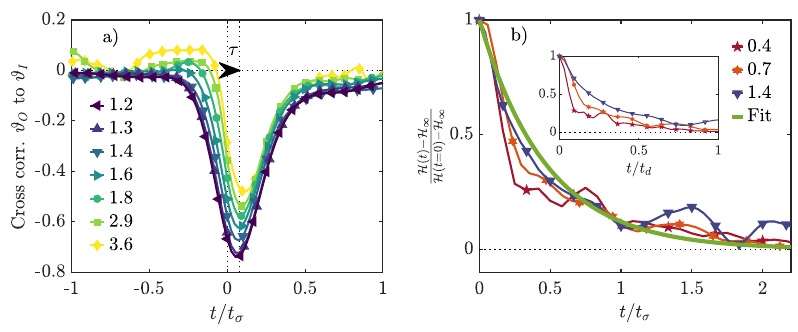}
    \caption{a) Cross-correlation between outer and inner strain. The negative peak ranges from $-0.75$ at $\mathrm{We}=1.2$, to $-0.5$ at $\mathrm{We}=3.6$, and is found at $\tau \approx 0.08~t_{\sigma}$. b) Ensemble-averaged surface energy relaxation of ellipsoidal drops at different $\mathrm{We}$ (as indicated in the legend), fitted with $\exp \left(-c_d t/t_{\sigma} \right)$, with $c_{d}=3.5$. The inset reproduces the relaxation behaviour but normalised with the eddy-turnover time $t_{d}$.}
    \label{fig:fig3}
\end{figure}
The time delay between the two strains is quantified by their temporal cross-correlation in figure~\ref{fig:fig3}a. The temporal cross-correlation exhibits a clear negative peak at $t \approx \tau$, which suggests that the outer strain precedes the inner strain, i.e. the inner strain can be considered as a passive, delayed response opposing interfacial stretching by outer eddies. We find that the peak increases in absolute magnitude as the Weber number decreases: the inner flow dynamics increasingly follow the dynamics of the outer flow. Moreover, $\tau$ exhibits a good collapse when normalised by the capillary time, with $\tau \approx 0.08\,t_{\sigma}$, corresponding to approximately one third of the autocorrelation time of the inner strain.

The following picture emerges: outer eddies that are independent of the interface dynamics generate stretching motions of duration $ \sim t_{\vartheta_{O}}$ at the interface and increase the drop's surface energy. After a brief delay, $\tau$, these stretching motions are quickly opposed by interfacial forces and the drop relaxes by injecting its excess energy back into the velocity field through the genesis of turbulent motions in the drop and its surrounding boundary layer (inner eddies). As a consequence, the outer strain term in equation~\eqref{eq:master} can be modelled as an external stochastic forcing agnostic to the drop dynamics, similar to the external forcing used by \citet{riviere2024bubble} in their harmonic oscillator model, but characterised by the p.d.f.\ in figure~\ref{fig:fig2}a and the autocorrelation function in figure~\ref{fig:fig2}c. By contrast, the inner strain term is a response to the outer strain and opposes it by removing energy from the interface at a rate dictated by the capillary time. This suggests that the relaxation of the drop toward equilibrium may be modelled as a damping term in the equation for the surface energy, equation~\eqref{eq:master}, as customary in harmonic oscillator models of bubble breakup \citep{risso1998oscillations,riviere2024bubble}. 

\subsection{Modelling the inner strain as a damping term in the equation for the surface energy}

We performed DNS in which an ellipsoidal drop was inserted into the turbulent flow. These simulations were initialised in the same way as the simulations with a spherical drop but by placing the major axis of the prolate ellipsoid aligned with one of the periodic directions. For each $\mathrm{We}=0.4, 0.7, 1.4$, we performed $32$ independent runs to examine the relaxation of the drop toward its equilibrium surface energy, $\mathcal{H}_{\infty}$, as excess energy is transferred to inner eddies. The dependence of $\mathcal{H}_{\infty}$ on the Weber number is shown in appendix~\ref{ap:Hinf}. In figure~\ref{fig:fig3}b, we show the ensemble-averaged surface energy as a function of time. The lines collapse when scaled with $t_{\sigma}$ and exhibit an approximate exponential decay toward equilibrium, as expected from a linearly damped motion. We obtained the damping rate, $c_d$, by fitting the three curves in figure~\ref{fig:fig3}b as
\begin{align}
\frac{\mathcal{H}\left(t\right) - \mathcal{H}_{\infty}(\mathrm{We})}{\mathcal{H}\left(0\right)-\mathcal{H}_{\infty}(\mathrm{We})} \approx \exp \left( - c_{d}\frac{t}{t_{\sigma}}\right) \text{.}
\label{eq:dampfit}
\end{align}
We found that a Weber-independent value of $c_{d} \approx 3.5$ fits the relaxation process well (green thick line in figure~\ref{fig:fig3}b). We note that this value is similar to the inverse of the inner strain's autocorrelation time $1/t_{\vartheta_{I}} \approx 4.5$. Although this agreement might be coincidental, it is physically meaningful: $c_d$ is the rate at which the inner flow extracts energy from the interface, whereas the autocorrelation time characterises the time scale over which inner eddies are generated by interfacial compression.

These results support that the inner strain term in equation~\eqref{eq:master} can be modelled as a linear damping,
\begin{equation}\label{eq:damp}
\vartheta_I \approx - \frac{c_d}{ t_\sigma} \left(\mathcal{H}-\mathcal{H}_{0}\right).
\end{equation}
Note that $\mathcal{H}_{0}$ enters the damping term (instead of $\mathcal{H}_{\infty}$) because, in the absence of outer stretching ($\vartheta_O=0$ in equation~\eqref{eq:master}), the drop relaxes to a spherical shape. The equilibrium value $\mathcal{H}_{\infty}$ is the solution of the time-average of equation~\eqref{eq:master}, in which the damping term exactly balances the average outer stretching, see appendix~\ref{ap:Hinf}. If $\mathcal{H}_{\infty}$ were used in equation~\eqref{eq:damp} instead of $\mathcal{H}_{0}$, a bias toward stronger deformations would be introduced. In the runs initialised with ellipsoidal drops, the outer stretching is only negligible in the early stages and thus in the fit (equation~\eqref{eq:dampfit}) the equilibrium surface energy must be used as reference for consistency.

 \citet{risso1998oscillations} described bubble deformation with a harmonic oscillator model for the bubble's second mode of vibration. They included the effect of the turbulent flow as an external forcing and considered damping according to the classical theory of \citet{lamb1932hydrodynamics} for the relaxation of slightly deformed drops and bubbles in a quiescent fluid. In Lamb's theory, the damping of small-amplitude interfacial oscillations is assumed to be determined by the viscosity of the continuous phase. \citet{miller1968oscillations} and \citet{Prosperetti1980} developed more general linear theories to account for viscous effects in both phases, and for the generation and diffusion of vorticity at the interface. However, \citet{riviere2024bubble} showed that, in turbulence, the effective damping of bubble shape oscillations is approximately one order of magnitude larger than Lamb's prediction. They attributed this excess to a turbulent boundary layer of thickness $\approx 0.1d$ surrounding the bubble and modelled it with an eddy viscosity based on inertial eddies, i.e. with a damping rate proportional to $t_d^{-1}$. The damping considered in our model represents a similar effective mechanism: the energy extracted from the deforming interface by the inner flow, i.e. by the vortical motions (inner eddies) inside the drop and in its surrounding boundary layer, up to a distance $\Delta=0.19d$ from the interface (of the same order as the boundary layer identified by \citet{riviere2024bubble}). This energy is eventually dissipated by viscosity at small scales, but the damping term itself models the energy extraction from the interface, and not the molecular viscous dissipation of interfacial oscillations. In contrast with the eddy-viscosity scaling of \citet{riviere2024bubble}, we show in the inset of figure~\ref{fig:fig3}b that the measured relaxation does not collapse well when scaled with the eddy-turnover time $t_d$. It appears that at low $\mathrm{We}\lesssim 2$, the flow around the drop is rapidly stirred by its oscillatory motion, with a characteristic velocity $d/t_\sigma$. This suggests an effective kinematic viscosity of order $d^2/t_\sigma$, consistent with the capillary-time scaling observed by \citet{roa2023droplet} for the relaxation of highly deformed drops at $\mathrm{We}=0.9$. 

We emphasise that the present damping model of the inner flow is restricted to cases with matched density and viscosity and does not necessarily capture the dynamics at different density and viscosity ratios (as in the case of bubbles). Consequently, the parameter $c_d$ must be adjusted to correctly account for the properties of the inner flow, particularly the density and viscosity ratios. Moreover, for unequal densities, $t_\sigma$ should account for the inertia of both phases \citep{lamb1932hydrodynamics}.

\section{Model of drop deformation and equation for the breakup rate}\label{sec:model}

By substituting equation~\eqref{eq:damp} into equation~\eqref{eq:master}, after scaling time with $t_d$ and energies with $\mathcal{H}_{0}$, we arrive at the following dimensionless stochastic differential equation for the excess surface energy 
 \begin{equation}\label{eq:model}
\frac{\mathrm{d} \xi}{\mathrm{d} \hat{t}}=\frac{\mathrm{We}}{\pi} \, \hat \vartheta_O  - \frac{c_d}{\sqrt{\mathrm{We}}} \xi,
\end{equation}
where $\hat \vartheta_O=\vartheta_O/(\rho \varepsilon d^3)$ is the dimensionless outer strain. The model can be interpreted as follows. Outer eddies stretch the interface and increase the surface energy of the drop. The stretching amplitude and statistics, given by $\hat \vartheta_O$, are solely a property of the turbulent flow. However, the power input is proportional to the Weber number, i.e. a fixed strain rate increases the surface energy at a rate proportional to the Weber number. The stretching by outer eddies is opposed by the conversion of surface energy into kinetic energy through the generation of inner eddies. This energy conversion acts as a sink in the equation, modelled here as a linear damping. The lower the Weber number, the faster the damping. We stress that while equation~\eqref{eq:master} is exact, the model equation~\eqref{eq:model} is not, because it reduces the inner dynamics to an energy sink. Note also that the damping response should be delayed by $\tau$. This delay is very small and is neglected for the sake of simplicity. 

\subsection{Solution of the model equation in the low-$\mathrm{We}$ limit}\label{sec:lowWe_sol}

Equation~\eqref{eq:model} can be integrated numerically by simulating  $\hat \vartheta_O$ as a stochastic process with dimensionless autocorrelation time $\hat t_{\vartheta_{O}}\approx 0.65$ and the probability density function shown in figure~\ref{fig:fig2}a. However, in the low-$\mathrm{We}$ limit it admits a simple analytical solution, which is obtained in what follows. 

The key consideration is that cumulative outer stretching effects may only occur if the damping time is longer than the typical duration of an outer-stretching event, given by the autocorrelation time. Their ratio, $\sqrt{\mathrm{We}}/(c_d\hat t_{\vartheta_{O}})\approx 0.44 \sqrt{\mathrm{We}}$, implies that at low $\mathrm{We}$ the energy injected by an event of outer stretching is rapidly damped before the next event arrives. We thus solve equation~\eqref{eq:model} for the case of a single, isolated outer-stretching event. Specifically, we consider an event of constant intensity $\hat{\vartheta}^e_O \rho \varepsilon d^3$ and duration $\hat{t}_{\vartheta_{O}}$. In this case, the integration of equation~\eqref{eq:model} from $\hat t=0$ to $\hat t=\hat{t}_{\vartheta_{O}}$ yields
\begin{equation}\label{eq:H1}
\xi \left(\hat{t}_{\vartheta_{O}}\right)= \frac{\mathrm{We}^{3/2}}{c_d\,\pi} \hat{\vartheta}^e_O\left(1-\delta\right),
\end{equation}
where 
\begin{equation}\label{eq:delta}
\delta=e^{-c_d \, \hat{t}_{\vartheta_{O}}\,\mathrm{We}^{-1/2}}. 
\end{equation}

\subsection{Breakup criterion}

\begin{figure}
    \centering
    \includegraphics[width=\textwidth]{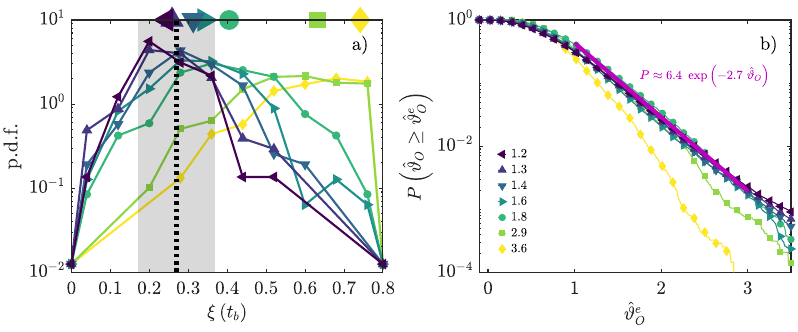}
    \caption{a) Statistics of the excess surface energy $\xi$, equation~\eqref{eq:xi}, at breakup time $t=t_{b}$. The averaged surface energy at breakup is represented at the top of the plot with large markers. When plotting the binned statistics, excess surface energies $\xi \geq 0.8$ are set equal to $\xi = 0.8$ for clarity. The vertical line denotes $\xi = 0.27$ and the shaded region plus/minus one standard deviation. b) Complementary c.d.f. of outer-stretching events at different Weber numbers (as indicated in the legend). In purple, an exponential fit to the complementary c.d.f.\ The fit is restricted to cases with $\mathrm{We} < 2$ and to the range $2 \leq \hat{\vartheta}_{O}^{e} \leq 2.5$.}
    \label{fig:fig4}
\end{figure}

Equation~\eqref{eq:H1} models the increase in excess surface energy, with respect to a spherical drop, as the result of a single, isolated outer-stretching event, whose intensity $\hat{\vartheta}^{e}_{O}$ is sampled from the p.d.f.\ in figure~\ref{fig:fig2}a. Following \citet{andersson2006breakup}, we assume that breakup occurs (on average) when $\xi(t_b) \gtrsim \xi_{c}=0.27$, at which point the surface energy exceeds that of two identical spherical drops of the same total volume as the initial drop. Although there is clearly no single $\xi_{c}$ above which all drops break (figure~\ref{fig:fig4}a), there is a shift towards a critical average breakup energy of $\xi_c \approx 0.27$ as the Weber number decreases. This is consistent with the measured energy at breakup in experiments \citep{andersson2006breakup} and DNS \citep{vashisth2025dynamics}. In appendix~\ref{ap:Hinf}, we show that the increase in $\xi(t_b)$ that occurs as $\mathrm{We}$ increases is largely compensated by the increase in equilibrium energy $\mathcal{H}_\infty$ with $\mathrm{We}$, so that the increment in energy necessary for breakup with respect to the equilibrium energy remains constant and close to $0.27$. Setting a deterministic threshold for breakup is one of the main limitations of our model. In fact, the definition of the conditions and time at which breakup is inevitable are the subject of active research \citep{qi2024breaking,roa2026critical,cannon2026strain} and require a probabilistic approach (discussed in \S\ref{sec:Conclusion}).

%We argue that this is one of the main limitations of our model..

\subsection{Extreme-event approach to predict breakup}\label{sec:extreme}

The large damping at low $\mathrm{We}$ implies that breakup must primarily occur as a result of a single, extreme outer-stretching event, i.e. a rare event of intense energy injection represented by the tail of the p.d.f.\ of figure~\ref{fig:fig2}a. Further evidence for this hypothesis is provided in appendix~\ref{ap:Conditional}, where we show that (i) for $\mathrm{We}<2$ breakup is systematically preceded by intense outer-stretching events, whose magnitude increases as the Weber number decreases; (ii) these events are rare, with their arrival times described by a Poisson process. The increase in excess surface energy caused by a single outer-stretching event, equation~\eqref{eq:H1}, satisfies the breakup condition, $\xi>\xi_c=0.27$,  provided that
\begin{align}
    \hat{\vartheta}^e_O \gtrsim \dfrac{\pi \, \xi_c  \,c_d}{1-\delta} \mathrm{We}^{-3/2} \text{.}
\end{align}
The probability of breakup can then be estimated from the p.d.f.\ $\left( p \right)$ of $\hat \vartheta_{O}$ as
\begin{align}
    P_{b} = P \left( \hat{\vartheta}_{O}  \geq \hat{\vartheta}^e_O \right) = \int_{\hat{\vartheta}^e_O}^{\infty} p~\mathrm{d} \hat{\vartheta}_{O} \text{,}
\end{align}
where $p$ has a long exponential tail (figure~\ref{fig:fig2}a). Hence, the probability of an event of sufficiently large intensity is
\begin{equation}\label{eq:tail}
P_b \approx \alpha e^{-\gamma \hat{\vartheta}^{e}_{O}}, 
\end{equation}
with $\alpha\approx 6.4$ (figure~\ref{fig:fig4}b). This can be written as  
\begin{equation}\label{eq:pb}
P_b \approx \alpha e^{-\chi \, \mathrm{We}^{-3/2}},
\end{equation}
where 
\begin{equation}\label{eq:chi}
\chi=\gamma\,\pi \, \xi_c  \,c_d/(1-\delta)\text{.} 
\end{equation}
The breakup rate is the frequency of independent outer-stretching events, given by the inverse of the autocorrelation time,  $1/\hat t_{\vartheta_{O}}$, times the probability that an event exceeds the breakup threshold,
\begin{equation}\label{eq:br}
\hat{\kappa} = \frac{\alpha}{\hat t_{\vartheta_{O}}} \, e^{-\chi \, \mathrm{We}^{-3/2}}.
\end{equation}
This equation is a prediction of the breakup rate, which is obtained using only data from DNS concerning (i) the outer turbulent flow ($\alpha$,  $\hat t_{\vartheta_{O}}$ and $\gamma$); (ii) the relaxation of ellipsoidal drops in turbulent flow ($c_d$) and the constant breakup threshold ($\xi_c$). All the parameters of the model are listed in table~\ref{tab:chi}, together with the uncertainty in their determination. In the next section, equation~\eqref{eq:br} is compared to the breakup rates obtained from the ensembles of DNS from \citet{AlbertoSciAdv} and those performed in this paper (table~\ref{tab:Database}).

\begin{table}
\centering
\caption{Parameters obtained from the DNS, with their corresponding standard deviations. For the calculation of the parameters we excluded the cases with $\mathrm{We}>2$. As $\mathrm{We}$ increases, breakup ceases to be solely due to extreme events and cumulative effects arise. In addition, the temporal cross-correlation of outer to inner eddies degrades and the inner strain progressively contributes to deformation, thereby reducing the validity of the underlying low-$\mathrm{We}$ assumption.}
\label{tab:chi}
\begin{tabular}{c|ccccc}
\hline
Parameter & $\gamma$ & $\xi_{c}$ & $c_{d}$ & $\hat{t}_{\vartheta_{O}}$ & $\alpha$\\ \hline
Mean value & 2.68 & 0.27 & 3.5 & 0.65 & 6.4 \\
Standard deviation & 0.17 & 0.1 & 0.2 & 0.03 & 1.3
\end{tabular}
\end{table}

\subsection{Breakup rate and breakup time}\label{sec:dns}

\begin{figure}
    \centering
    \includegraphics[width=\textwidth]{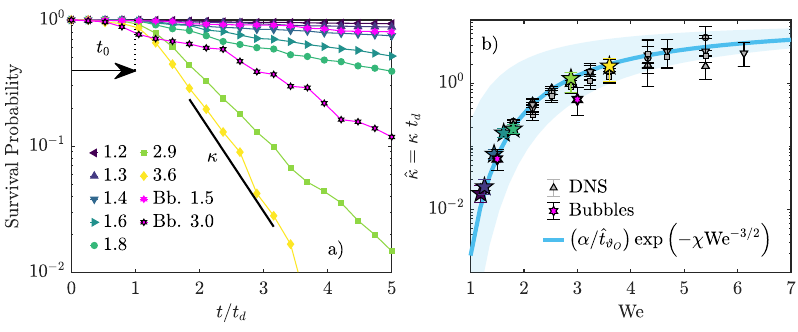}
    \caption{a) Survival probability obtained from DNS of drops at different Weber numbers (with the same colours and markers as in previous figures) and with magenta hexagons the bubble breakup data extracted from the DNS of \citet{perrard2021bubble} at two Weber numbers. b) Breakup rate of drops at different Weber numbers. Grey markers denote data measured in the numerical simulations performed by \citet{AlbertoSciAdv} and big coloured markers the ones carried out here. The thick blue solid line corresponds to equation~\eqref{eq:br}, where the slope $\chi = \gamma\,\pi \, \xi_c  \,c_d \, /\left(1 - \delta \right)$, $\alpha \approx 6.4$ and $\hat{t}_{\vartheta_{O}} \approx 0.65$ are obtained from the DNS results, table~\ref{tab:chi}. The shaded region covers the 10$^\text{th}$ to 90$^\text{th}$ percentile interval obtained from Monte Carlo uncertainty propagation, where the parameters were sampled according to the standard deviations reported in table~\ref{tab:chi}.}
    \label{fig:fig5}
\end{figure}

The methodology used to extract the breakup rate from ensembles of the DNS is detailed by \citet{AlbertoSciAdv}. It involves two key aspects. First, the breakup rate is well defined only in a statistically steady state, but as shown in appendix~\ref{ap:Hinf}, in the DNS, the drop requires a time ($t\lesssim t_d$) to equilibrate with the ambient turbulence. Second, there is a delay between the time at which a critically deformed state is reached and the breakup time \citep{qi2024breaking,roa2026critical,cannon2026strain}. This delay is of order $t_\sigma$ and thus negligible at low $\mathrm{We}$, but becomes substantial as $\mathrm{We}$ increases. The initial equilibration time and the rupture delay are not part of the model and must hence be excluded in the estimation of the breakup rate from DNS ensembles \citep{AlbertoSciAdv}. In each individual DNS, these times cannot be determined; however, their combined effect (termed $t_0$) is visible in the survival probability (figure~\ref{fig:fig5}a), where a statistically steady state (corresponding to a constant slope of the survival probability in logarithmic scale) is reached for $t\gtrsim t_0\in[0,2]t_d$. This state is characterised by a memoryless process with an exponential tail and thus a constant breakup rate. We stress that it is crucial to account for $t_0$ and its uncertainty in the analysis; otherwise, the breakup rate depends on the observation time. Following \citet{AlbertoSciAdv}, $t_0$ is chosen sufficiently large for the estimated breakup rate $\kappa$ to become independent of its value, with the residual uncertainty due to its selection included in the error bars of $\kappa$. More details on how we compute $t_{0}$ and $\kappa$ can be found in \citet{AlbertoSciAdv}, in particular figure S2 of their supplementary material. %Second, DNS are customarily terminated after a cutoff time, $t_c$, even if the particle has not broken yet. This truncation of the data must be accounted for in the maximum likelihood estimation of $\kappa$ \citep{lawless}. 

The breakup rates estimated from our ensembles of DNS are shown as coloured symbols in figure~\ref{fig:fig5}b, where the error bars contain both the statistical (sampling) uncertainty, and the uncertainty in determining $t_0$. The agreement with the model prediction, equation~\eqref{eq:br}, is entirely within the propagated uncertainty in the model constants. In addition, we analysed the bubble breakup times from the DNS of \citet{perrard2021bubble} (magenta hexagons in figure~\ref{fig:fig5}a), and found that the corresponding breakup rate is well predicted by our model at $\mathrm{We}\lesssim 3.5$, figure~\ref{fig:fig5}b. \\
Indeed, \citet{riviere2024bubble} have shown that at low Weber numbers, the outer turbulent flow drives bubble deformation. They also showed that enhanced damping occurs mainly in the boundary layer of liquid adjacent to the bubble, which is included in our inner region. A more thorough investigation of bubble breakup is warranted, but the present results are encouraging.

\subsection{Mean breakup time}
\begin{figure}
    \centering
    \includegraphics[width=\textwidth]{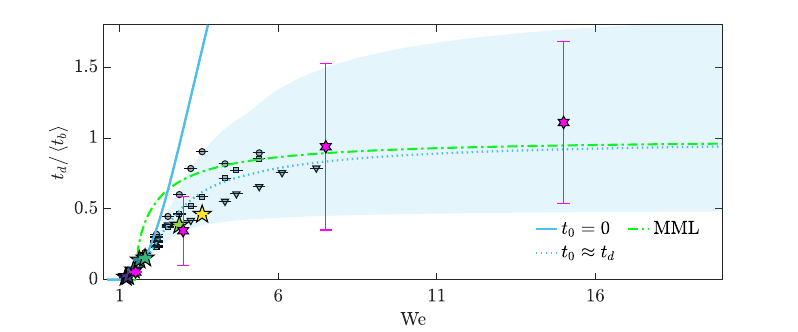}
    \caption{Average breakup time $\langle t_{b}\rangle$ from our model (solid line) and adjusted with $t_{0}=t_{d}$ (dotted line). The shaded region covers adjustments to the model with initial times $t_{0} \in \left[0.5~ t_{d},2~t_{d} \right]$. The dashed-dotted green line (MML) shows the fit proposed by \citet{martinez1999breakup, martinez2010considerations} for bubble breakup: $t_{d}/\left \langle t_{b} \right \rangle \approx C_{g} \sqrt{1-\frac{\mathrm{We}_{c}}{\mathrm{We}}}$, with $C_{g}=1.2$ and $\mathrm{We}_{c}=1.5$ (from a least-squares fit). The symbols are as in figure~\ref{fig:fig5}b.}
    \label{fig:fig6}
\end{figure}
 
In most previous studies of drop and bubble breakup, $t_0$ was not considered in the analysis and the breakup rate was computed as the inverse of the mean breakup time of the sample, $\left \langle t_{b} \right \rangle$. Approximating the mean requires much smaller sample sizes than the true breakup rate $\kappa$, which additionally requires estimating $t_0$ and then discarding all breakups with $t_b<t_0$. However, $\left \langle t_{b} \right \rangle$ is highly influenced by the way in which drops/bubbles are initiated in the DNS or experiments and accordingly presents much more scatter (see figure~\ref{fig:fig6}, where the same ensembles of data as used in figure~\ref{fig:fig5} to estimate $\kappa$ have been used to compute the sample mean). At low $\mathrm{We}$, $t_0\ll \kappa^{-1}$ and this influence is small, because breakup times are long. Accordingly, $\langle t_{b} \rangle \approx \kappa^{-1}$ and our model for the breakup rate yields excellent agreement with the mean breakup time. Introducing $t_0$ into the model, i.e. $\langle t_b \rangle =t_0+\kappa^{-1}$ (for simplicity, $t_0=t_d$ was chosen), reasonably approximates the mean breakup time data at all Weber numbers and exhibits a trend similar to that of the empirical model proposed to approximate the mean breakup times of bubbles in experiments of jet flows \citep{martinez1999breakup, martinez2010considerations}. However, we stress that, unlike the memoryless breakup rate $\kappa$ the mean breakup time is not an intrinsic property of drop breakup in HIT and strongly depends on the initial conditions.

\section{Conclusion}\label{sec:Conclusion}

We have derived a simple model that quantitatively links properties of the turbulent flow to the fragmentation rate of drops at low Weber numbers. This connection has been explored in many previous studies \citep{coulaloglou1977description,baldyga1998drop,qi2022fragmentation,riviere2024bubble,ni2024deformation,Wu2026} and was formalised here by applying the decomposition of interfacial strain generated by outer and inner eddies introduced by \citet{AlbertoJFM} to the equation for the surface energy. Outer eddies are not affected by the drop and drive deformation, whereas inner eddies are generated by capillary forces and (on average) act so the drop relaxes. This relaxation occurs at the capillary time scale. Since energy is injected by the outer eddies at the eddy-turnover time scale, the Weber number controls the competition between the two processes. Two distinct limits emerge. At large $\mathrm{We}$ (or low damping), the scaling of \citet{coulaloglou1977description} is recovered, in which $\mathrm{We}^{-1}$ enters the exponential dependence of the breakup rate (see appendix~\ref{ap:CT77}). At small $\mathrm{We}$, damping dominates and the breakup rate decreases more rapidly, with $\mathrm{We}^{-3/2}$ in the exponent. Equation~\eqref{eq:br} smoothly connects both scalings through the function $\delta$. Although our model is slightly more complicated than the equation proposed by \citet{coulaloglou1977description}, it was obtained step by step from the governing equations (stating all assumptions and justifying them with DNS data). Further, it has no parameter fitted directly to the breakup rates, and thus a higher predictive power. All constants in the model can be measured a priori and quantify properties of the outer turbulent flow and the drop response.

The least justified assumption in our model is that breakup occurs on average at a critical excess surface energy, $\xi(t)\geq\xi_{c}=0.27$. We have shown that the average energy at breakup time, $\mathcal{H}\left( t_{b} \right)$, increases with the Weber number. However, the drop relaxes to a quasi-equilibrium surface energy, $\mathcal{H}_\infty$, that also increases with the Weber number. On the whole, the difference between $\mathcal{H}_{\infty}/\mathcal{H}_0$ and $\mathcal{H}\left( t_{b} \right)/\mathcal{H}_0$ (i.e. the increment necessary for breakup) remains close to $\xi_{c}=0.27$ at all the Weber numbers considered. More importantly, the physically relevant breakup threshold should not be extracted at the moment of breakup, but rather at the time when breakup becomes irreversible, when the so-called critically deformed state is reached \citep{haakansson2022criterion,vashisth2025dynamics,roa2026critical,cannon2026strain}. The difficulty here is that accurately estimating the critically deformed state would necessitate a probabilistic approach, as recently implemented to study the predictability of extreme dissipation events in Kolmogorov flow by \citet{vela2024large}, and turbulence decay in pipe flow by \citet{montesdeoca2025probabilistic}. In these works, massive ensembles of simulations were used to quantify how the predictability of an event (here breakup) degrades when one goes back in time. The last point in time at which breakup cannot be predicted with certainty would correspond to the critically deformed state. We expect this time to depend on the specific realisation, as shown for turbulence decay in pipe flow \citep{montesdeoca2025probabilistic}. We thus argue that for the prediction of the (average) breakup rate proposed in this paper, an average critical excess surface energy sits at a sweet spot between simplicity and accuracy.

We stress that our prediction is only applicable for viscosity and density ratios close to $1$. Indeed, \citet{AlbertoSciAdv} showed that for viscosity ratios between $0.5$ and $2$ the breakup rate is unaffected (at the Reynolds number considered in this paper). For sufficiently viscous drops, internal viscous time scales enhance damping forces, eventually dominating the breakup dynamics \citep{Farsoiya_Liu_Daiss_Fox_Deike_2023}.
We suggest that the model can be adapted with a viscosity-dependent damping rate, which can be estimated with DNS initialised with ellipsoidal drops. Regarding bubble fragmentation, the good agreement of our model with the bubble DNS of \citet{perrard2021bubble} and the similar thickness of the boundary layer surrounding drops \citep{AlbertoJFM} and bubbles \citep{riviere2024bubble} suggest that the damping mechanism may be similar. These suggestions merit scrutiny in future studies. We further speculate that our model  can also be extended to more complex flow configurations \citep{lasheras2002review}, non-Newtonian fluids \citep{shimizu1999drop}, surfactant-laden flows \citep{cannon24morphology}, and particularly high-Reynolds-number turbulence \citep{ni2024deformation}. In all these cases, the forcing experienced by the particle should remain primarily governed by the stretching of the outer, drop-independent field. Extending the model would then require replacing the tail of the c.d.f.\ in figure~\ref{fig:fig4}b with the appropriate distribution, correctly capturing the intensity of extreme stretching by outer eddies \citep{buaria2019extreme}. 

In summary, we expect our model to remain valid beyond the conditions studied here through appropriate scalings of the outer forcing and the energy extraction by the inner flow. Both can be measured a priori with DNS. The good match between our model and bubble breakup data shown in this paper is encouraging in this respect.

%%%%%%%%%%%%%%%%%%%%%%%%%%%%%%%%%%%%%%%%%%%%%%%%%%%%%%%%%%%%%%%%%%%%%%%%%
% BACKSECTIONS
%%%%%%%%%%%%%%%%%%%%%%%%%%%%%%%%%%%%%%%%%%%%%%%%%%%%%%%%%%%%%%%%%%%%%%%%%
%\backsection[Supplementary data]{}
\backsection[Acknowledgements]{We are grateful to Ali\'enor Rivi\`ere for fruitful and insightful discussions and sharing her bubble breakup data with us.}

\backsection[Funding]{This work was funded by the Deutsche Forschungsgemeinschaft (DFG, German Research Foundation) under Germany's Excellence Strategy – EXC-3036 The Martian Mindset, project number: 533607631. This work was also supported by the European Union's Horizon Europe research and innovation programme under the Marie Sklodowska-Curie Actions (MSCA), grant agreement No.~101281052 (RIACO). Part of this work was carried out during the Sixth Madrid Turbulence Workshop hosted by Prof.~Javier Jim\'{e}nez, and supported by the European Research Council (ERC) under the Advanced Grant \emph{CAUST} (ERC-AdG-101018287). They are here gratefully acknowledged.}

\backsection[Declaration of interests]{The authors declare no conflict of interest.}

\backsection[Data availability statement]{All data and source code required to reproduce the results of this study are openly available in \href{https://www.pangaea.de/}{PANGAEA} at \href{https://doi.pangaea.de/10.1594/PANGAEA.994057}{https://doi.pangaea.de/10.1594/PANGAEA.994057}.}

\bibliographystyle{jfm}
\bibliography{jfm}

\begin{thebibliography}{58}
\expandafter\ifx\csname natexlab\endcsname\relax\def\natexlab#1{#1}\fi
\def\au#1{#1} \def\ed#1{#1} \def\yr#1{#1}\def\at#1{#1}\def\jt#1{\textit{#1}}
  \def\bt#1{#1}\def\bvol#1{\textbf{#1}} \def\vol#1{#1} \def\pg#1{#1}
  \def\publ#1{#1}\def\arxiv#1{#1}\def\org#1{#1}\def\st#1{\textit{#1}}

\bibitem[Andersson \& Andersson(2006)]{andersson2006breakup}
{\sc \au{Andersson, Ronnie} \& \au{Andersson, Bengt}} \yr{2006}  \at{On the
  breakup of fluid particles in turbulent flows}.  \jt{AIChE Journal}
  \bvol{52}~(6),  \pg{2020--2030}.

\bibitem[Baldyga \& Podg{\'o}rska(1998)]{baldyga1998drop}
{\sc \au{Baldyga, Jerzy} \& \au{Podg{\'o}rska, Wioletta}} \yr{1998}  \at{Drop
  break-up in intermittent turbulence: Maximum stable and transient sizes of
  drops}.  \jt{The Canadian Journal of Chemical Engineering}  \bvol{76}~(3),
  \pg{456--470}.

\bibitem[Beckedorff {\em et~al.\/}(2025)Beckedorff, Caridi \&
  Soldati]{beckedorff2025jet}
{\sc \au{Beckedorff, Leonel}, \au{Caridi, Giuseppe~CA} \& \au{Soldati,
  Alfredo}} \yr{2025}  \at{Jet-stirred homogeneous isotropic turbulent water
  tank for bubble and droplet fragmentation}.  \jt{Review of Scientific
  Instruments}  \bvol{96}~(2).

\bibitem[Brizzolara {\em et~al.\/}(2024)Brizzolara, Naudascher, Rosti, Stocker,
  Boffetta, Mazzino \& Holzner]{brizzolara2024immiscible}
{\sc \au{Brizzolara, Stefano}, \au{Naudascher, Robert}, \au{Rosti,
  Marco~Edoardo}, \au{Stocker, Roman}, \au{Boffetta, Guido}, \au{Mazzino,
  Andrea} \& \au{Holzner, Markus}} \yr{2024}  \at{Immiscible {Rayleigh--Taylor
  turbulence: Implications} for bacterial degradation in oil spills}.
  \jt{Proceedings of the National Academy of Sciences}  \bvol{121}~(11),
  \pg{e2311798121}.

\bibitem[Buaria {\em et~al.\/}(2019)Buaria, Pumir, Bodenschatz \&
  Yeung]{buaria2019extreme}
{\sc \au{Buaria, Dhawal}, \au{Pumir, Alain}, \au{Bodenschatz, Eberhard} \&
  \au{Yeung, Pui-Keun}} \yr{2019}  \at{Extreme velocity gradients in turbulent
  flows}.  \jt{New Journal of Physics}  \bvol{21}~(4),  \pg{043004}.

\bibitem[Cannon {\em et~al.\/}(2026)Cannon, Mor{\'o}n, Vela-Mart{\'\i}n \&
  Avila]{cannon2026strain}
{\sc \au{Cannon, Ianto}, \au{Mor{\'o}n, Daniel}, \au{Vela-Mart{\'\i}n, Alberto}
  \& \au{Avila, Marc}} \yr{2026} Strain-driven drop breakup in turbulence.
  \bt{In {\em Journal of Physics: Conference Series\/}}, ,  \vol{vol. 3230},
  \pg{p. 012006}. IOP Publishing.

\bibitem[Cannon {\em et~al.\/}(2024)Cannon, Soligo \&
  Rosti]{cannon24morphology}
{\sc \au{Cannon, Ianto}, \au{Soligo, Giovanni} \& \au{Rosti, Marco~E.}}
  \yr{2024}  \at{Morphology of clean and surfactant-laden droplets in
  homogeneous isotropic turbulence}.  \jt{Journal of Fluid Mechanics}
  \bvol{987},  \pg{A31}.

\bibitem[Cardesa {\em et~al.\/}(2017)Cardesa, Vela-Martín \&
  Jiménez]{cardesa}
{\sc \au{Cardesa, José~I.}, \au{Vela-Martín, Alberto} \& \au{Jiménez,
  Javier}} \yr{2017}  \at{The turbulent cascade in five dimensions}.
  \jt{Science}  \bvol{357}~(6353),  \pg{782--784},  \arxiv{arXiv:
  https://www.science.org/doi/pdf/10.1126/science.aan7933}.

\bibitem[Coulaloglou \& Tavlarides(1977)]{coulaloglou1977description}
{\sc \au{Coulaloglou, CA} \& \au{Tavlarides, Lawrence~L}} \yr{1977}
  \at{Description of interaction processes in agitated liquid-liquid
  dispersions}.  \jt{Chemical Engineering Science}  \bvol{32}~(11),
  \pg{1289--1297}.

\bibitem[Deane \& Stokes(2002)]{deane2002scale}
{\sc \au{Deane, Grant~B} \& \au{Stokes, M~Dale}} \yr{2002}  \at{Scale
  dependence of bubble creation mechanisms in breaking waves}.  \jt{Nature}
  \bvol{418}~(6900),  \pg{839--844}.

\bibitem[Deike(2022)]{deike2022mass}
{\sc \au{Deike, Luc}} \yr{2022}  \at{Mass transfer at the ocean--atmosphere
  interface: the role of wave breaking, droplets, and bubbles}.  \jt{Annual
  Review of Fluid Mechanics}  \bvol{54}~(1),  \pg{191--224}.

\bibitem[Dodd \& Ferrante(2016)]{DoddFerrante2016}
{\sc \au{Dodd, Michael~S.} \& \au{Ferrante, Antonino}} \yr{2016}  \at{On the
  interaction of taylor length scale size droplets and isotropic turbulence}.
  \jt{Journal of Fluid Mechanics}  \bvol{806},  \pg{356--412}.

\bibitem[Eastwood {\em et~al.\/}(2004)Eastwood, Armi \&
  Lasheras]{eastwood2004breakup}
{\sc \au{Eastwood, Craig~Douglas}, \au{Armi, L} \& \au{Lasheras, JC}} \yr{2004}
   \at{The breakup of immiscible fluids in turbulent flows}.  \jt{Journal of
  Fluid Mechanics}  \bvol{502},  \pg{309--333}.

\bibitem[Farsoiya {\em et~al.\/}(2023)Farsoiya, Liu, Daiss, Fox \&
  Deike]{Farsoiya_Liu_Daiss_Fox_Deike_2023}
{\sc \au{Farsoiya, Palas~Kumar}, \au{Liu, Zehua}, \au{Daiss, Andreas}, \au{Fox,
  Rodney~O.} \& \au{Deike, Luc}} \yr{2023}  \at{Role of viscosity in turbulent
  drop break-up}.  \jt{Journal of Fluid Mechanics}  \bvol{972},  \pg{A11}.

\bibitem[H{\aa}kansson(2019)]{haakansson2019emulsion}
{\sc \au{H{\aa}kansson, Andreas}} \yr{2019}  \at{Emulsion formation by
  homogenization: {Current} understanding and future perspectives}.
  \jt{{Annual Review of Food Science and Technology}}  \bvol{10}~(1),
  \pg{239--258}.

\bibitem[H{\aa}kansson(2020)]{haakansson2020validity}
{\sc \au{H{\aa}kansson, Andreas}} \yr{2020}  \at{On the validity of different
  methods to estimate breakup frequency from single drop experiments}.
  \jt{Chemical Engineering Science}  \bvol{227},  \pg{115908}.

\bibitem[H{\aa}kansson {\em et~al.\/}(2022)H{\aa}kansson, Crialesi-Esposito,
  Nilsson \& Brandt]{haakansson2022criterion}
{\sc \au{H{\aa}kansson, Andreas}, \au{Crialesi-Esposito, Marco}, \au{Nilsson,
  Lars} \& \au{Brandt, Luca}} \yr{2022}  \at{A criterion for when an emulsion
  drop undergoing turbulent deformation has reached a critically deformed
  state}.  \jt{Colloids and Surfaces A: Physicochemical and Engineering
  Aspects}  \bvol{648},  \pg{129213}.

\bibitem[Hamlington {\em et~al.\/}(2008)Hamlington, Schumacher \&
  Dahm]{hamlington2008local}
{\sc \au{Hamlington, Peter~E.}, \au{Schumacher, J{\"{o}}rg} \& \au{Dahm, Werner
  J.~A.}} \yr{2008}  \at{Local and nonlocal strain rate fields and vorticity
  alignment in turbulent flows}.  \jt{Physical Review E}  \bvol{77}~(2),
  \pg{026303}.

\bibitem[Hinze(1955)]{hinze1955fundamentals}
{\sc \au{Hinze, Julius~O}} \yr{1955}  \at{Fundamentals of the hydrodynamic
  mechanism of splitting in dispersion processes}.  \jt{AIChE journal}
  \bvol{1}~(3),  \pg{289--295}.

\bibitem[Jacqmin(1999)]{jacqmin1999calculation}
{\sc \au{Jacqmin, David}} \yr{1999}  \at{Calculation of two-phase
  {Navier--Stokes} flows using phase-field modeling}.  \jt{Journal of
  computational physics}  \bvol{155}~(1),  \pg{96--127}.

\bibitem[Jiang {\em et~al.\/}(2010)Jiang, Siamas, Jagus \&
  Karayiannis]{jiang2010physical}
{\sc \au{Jiang, Xi}, \au{Siamas, George~A}, \au{Jagus, Krzysztof} \&
  \au{Karayiannis, Tassos~G}} \yr{2010}  \at{Physical modelling and advanced
  simulations of gas--liquid two-phase jet flows in atomization and sprays}.
  \jt{Progress in energy and combustion science}  \bvol{36}~(2),
  \pg{131--167}.

\bibitem[Jim{\'e}nez {\em et~al.\/}(1993)Jim{\'e}nez, Wray, Saffman \&
  Rogallo]{jimenez1993structure}
{\sc \au{Jim{\'e}nez, Javier}, \au{Wray, Alan~A}, \au{Saffman, Philip~G} \&
  \au{Rogallo, Robert~S}} \yr{1993}  \at{The structure of intense vorticity in
  isotropic turbulence}.  \jt{Journal of Fluid Mechanics}  \bvol{255},
  \pg{65--90}.

\bibitem[Kolmogorov(1949)]{kolmogorov1949disintegration}
{\sc \au{Kolmogorov, AN}} \yr{1949} On the disintegration of drops in a
  turbulent flow.  \bt{In {\em Dokl. Akad. Nauk SSSR\/}}, ,  \vol{vol.~66},
  \pg{p.~30}. sn.

\bibitem[Lamb(1932)]{lamb1932hydrodynamics}
{\sc \au{Lamb, Horace}} \yr{1932} {\em Hydrodynamics\/}, 6th edn.
  \publ{Cambridge: Cambridge University Press}.

\bibitem[Lasheras {\em et~al.\/}(2002)Lasheras, Eastwood, Mart{\i}nez-Baz{\'a}n
  \& Monta{\~n}es]{lasheras2002review}
{\sc \au{Lasheras, Juan~C}, \au{Eastwood, C}, \au{Mart{\i}nez-Baz{\'a}n, C} \&
  \au{Monta{\~n}es, Jose~L}} \yr{2002}  \at{A review of statistical models for
  the break-up of an immiscible fluid immersed into a fully developed turbulent
  flow}.  \jt{International Journal of Multiphase Flow}  \bvol{28}~(2),
  \pg{247--278}.

\bibitem[Maa\ss \& Kraume(2012)]{maass2012}
{\sc \au{Maa\ss, Sebastian} \& \au{Kraume, Matthias}} \yr{2012}
  \at{Determination of breakage rates using single drop experiments}.
  \jt{Chem.\ Eng.\ Sci.}  \bvol{70},  \pg{146 -- 164}.

\bibitem[Magaletti {\em et~al.\/}(2013)Magaletti, Picano, Chinappi, Marino \&
  Casciola]{magaletti2013sharp}
{\sc \au{Magaletti, Francesco}, \au{Picano, Francesco}, \au{Chinappi, Mauro},
  \au{Marino, Luca} \& \au{Casciola, Carlo~Massimo}} \yr{2013}  \at{The
  sharp-interface limit of the {Cahn--Hilliard/Navier--Stokes} model for binary
  fluids}.  \jt{Journal of Fluid Mechanics}  \bvol{714},  \pg{95--126}.

\bibitem[Mart{\'\i}nez-Baz{\'a}n {\em et~al.\/}(1999)Mart{\'\i}nez-Baz{\'a}n,
  Monta{\~n}es \& Lasheras]{martinez1999breakup}
{\sc \au{Mart{\'\i}nez-Baz{\'a}n, Carlos}, \au{Monta{\~n}es, Jose~L} \&
  \au{Lasheras, Juan~C}} \yr{1999}  \at{On the breakup of an air bubble
  injected into a fully developed turbulent flow. part 1. breakup frequency}.
  \jt{Journal of Fluid Mechanics}  \bvol{401},  \pg{157--182}.

\bibitem[Mart{\'\i}nez-Baz{\'a}n {\em et~al.\/}(2010)Mart{\'\i}nez-Baz{\'a}n,
  Rodr{\'\i}guez-Rodr{\'\i}guez, Deane, Monta{\~n}es \&
  Lasheras]{martinez2010considerations}
{\sc \au{Mart{\'\i}nez-Baz{\'a}n, Carlos}, \au{Rodr{\'\i}guez-Rodr{\'\i}guez,
  Javier}, \au{Deane, GB}, \au{Monta{\~n}es, Jose~L} \& \au{Lasheras, Juan~C}}
  \yr{2010}  \at{Considerations on bubble fragmentation models}.  \jt{Journal
  of Fluid Mechanics}  \bvol{661},  \pg{159--177}.

\bibitem[Mathijssen {\em et~al.\/}(2023)Mathijssen, Lisicki, Prakash \&
  Mossige]{mathijssen2023culinary}
{\sc \au{Mathijssen, Arnold~JTM}, \au{Lisicki, Maciej}, \au{Prakash, Vivek~N}
  \& \au{Mossige, Endre~JL}} \yr{2023}  \at{Culinary fluid mechanics and other
  currents in food science}.  \jt{Reviews of Modern Physics}  \bvol{95}~(2),
  \pg{025004}.

\bibitem[Miller \& Scriven(1968)]{miller1968oscillations}
{\sc \au{Miller, CA} \& \au{Scriven, LE}} \yr{1968}  \at{The oscillations of a
  fluid droplet immersed in another fluid}.  \jt{Journal of fluid mechanics}
  \bvol{32}~(3),  \pg{417--435}.

\bibitem[Montesdeoca {\em et~al.\/}(2025)Montesdeoca, Vela-Mart{\'\i}n \&
  Avila]{montesdeoca2025probabilistic}
{\sc \au{Montesdeoca, Daniel~Mor{\'o}n}, \au{Vela-Mart{\'\i}n, Alberto} \&
  \au{Avila, Marc}} \yr{2025}  \at{Probabilistic thresholds of turbulence decay
  in transitional shear flows}.  \jt{Journal of Fluid Mechanics}  \bvol{1022},
  \pg{A48}.

\bibitem[Mukherjee {\em et~al.\/}(2019)Mukherjee, Safdari, Shardt, Kenjereš \&
  Van~den Akker]{Mukherjee2019}
{\sc \au{Mukherjee, Siddhartha}, \au{Safdari, Arman}, \au{Shardt, Orest},
  \au{Kenjereš, Saša} \& \au{Van~den Akker, Harry E.~A.}} \yr{2019}
  \at{Droplet–turbulence interactions and quasi-equilibrium dynamics in
  turbulent emulsions}.  \jt{Journal of Fluid Mechanics}  \bvol{878},
  \pg{221–276}.

\bibitem[Ni(2024)]{ni2024deformation}
{\sc \au{Ni, Rui}} \yr{2024}  \at{Deformation and breakup of bubbles and drops
  in turbulence}.  \jt{Annual Review of Fluid Mechanics}  \bvol{56}~(1),
  \pg{319--347}.

\bibitem[Ohkitani \& Kishiba(1995)]{ohkitani1995nonlocal}
{\sc \au{Ohkitani, Koji} \& \au{Kishiba, Seigo}} \yr{1995}  \at{Nonlocal nature
  of vortex stretching in an inviscid fluid}.  \jt{Physics of Fluids}
  \bvol{7}~(2),  \pg{411--421}.

\bibitem[Perrard {\em et~al.\/}(2021)Perrard, Rivi{\`e}re, Mostert \&
  Deike]{perrard2021bubble}
{\sc \au{Perrard, St{\'e}phane}, \au{Rivi{\`e}re, Ali{\'e}nor}, \au{Mostert,
  Wouter} \& \au{Deike, Luc}} \yr{2021}  \at{Bubble deformation by a turbulent
  flow}.  \jt{Journal of Fluid Mechanics}  \bvol{920},  \pg{A15}.

\bibitem[Pope(1988)]{Pope1988}
{\sc \au{Pope, Stephen~B.}} \yr{1988}  \at{The evolution of surfaces in
  turbulence}.  \jt{International Journal of Engineering Science}
  \bvol{26}~(5),  \pg{445--469}.

\bibitem[Prosperetti(1980)]{Prosperetti1980}
{\sc \au{Prosperetti, Andrea}} \yr{1980}  \at{Free oscillations of drops and
  bubbles: the initial-value problem}.  \jt{Journal of Fluid Mechanics}
  \bvol{100}~(2),  \pg{333--347}.

\bibitem[Qi {\em et~al.\/}(2020)Qi, Masuk \& Ni]{qi2020towards}
{\sc \au{Qi, Yinghe}, \au{Masuk, Ashik Ullah~Mohammad} \& \au{Ni, Rui}}
  \yr{2020}  \at{Towards a model of bubble breakup in turbulence through
  experimental constraints}.  \jt{International Journal of Multiphase Flow}
  \bvol{132},  \pg{103397}.

\bibitem[Qi {\em et~al.\/}(2022)Qi, Tan, Corbitt, Urbanik, Salibindla \&
  Ni]{qi2022fragmentation}
{\sc \au{Qi, Yinghe}, \au{Tan, Shiyong}, \au{Corbitt, Noah}, \au{Urbanik,
  Carl}, \au{Salibindla, Ashwanth~KR} \& \au{Ni, Rui}} \yr{2022}
  \at{Fragmentation in turbulence by small eddies}.  \jt{Nature communications}
   \bvol{13}~(1),  \pg{469}.

\bibitem[Qi {\em et~al.\/}(2024)Qi, Xu, Tan, Zhong, Wu \& Ni]{qi2024breaking}
{\sc \au{Qi, Yinghe}, \au{Xu, Xu}, \au{Tan, Shiyong}, \au{Zhong, Shijie},
  \au{Wu, Qianwen} \& \au{Ni, Rui}} \yr{2024}  \at{Breaking bubbles across
  multiple time scales in turbulence}.  \jt{Journal of Fluid Mechanics}
  \bvol{983},  \pg{A24}.

\bibitem[Reid(1960)]{reid1960oscillations}
{\sc \au{Reid, William~Hill}} \yr{1960}  \bt{The oscillations of a viscous
  liquid drop}. {\em Tech. Rep.\/}.  \org{Division of Applied Mathematics,
  Brown University}.

\bibitem[Risso \& Fabre(1998)]{risso1998oscillations}
{\sc \au{Risso, Fr{\'e}d{\'e}ric} \& \au{Fabre, Jean}} \yr{1998}
  \at{Oscillations and breakup of a bubble immersed in a turbulent field}.
  \jt{Journal of Fluid Mechanics}  \bvol{372},  \pg{323--355}.

\bibitem[Rivi{\`e}re {\em et~al.\/}(2024)Rivi{\`e}re, Abahri \&
  Perrard]{riviere2024bubble}
{\sc \au{Rivi{\`e}re, Ali{\'e}nor}, \au{Abahri, Kamel} \& \au{Perrard,
  St{\'e}phane}} \yr{2024}  \at{Bubble shape oscillations in a turbulent
  environment}.  \jt{Journal of Fluid Mechanics}  \bvol{1001},  \pg{A26}.

\bibitem[Roa {\em et~al.\/}(2023)Roa, Renoult, Dumouchel \& Br{\"a}ndle~de
  Motta]{roa2023droplet}
{\sc \au{Roa, Ignacio}, \au{Renoult, Marie-Charlotte}, \au{Dumouchel,
  Christophe} \& \au{Br{\"a}ndle~de Motta, Jorge~C{\'e}sar}} \yr{2023}
  \at{Droplet oscillations in a turbulent flow}.  \jt{Frontiers in Physics}
  \bvol{11},  \pg{1173521}.

\bibitem[Roa {\em et~al.\/}(2026)Roa, Thiesset, Renoult, Dumouchel \&
  de~Motta]{roa2026critical}
{\sc \au{Roa, Ignacio}, \au{Thiesset, Fabien}, \au{Renoult, Marie-Charlotte},
  \au{Dumouchel, Christophe} \& \au{de~Motta, Jorge C{\'e}sar~Br{\"a}ndle}}
  \yr{2026}  \at{Critical deformation state of binary droplet breakup in
  turbulence}.  \jt{Journal of Fluid Mechanics}  \bvol{1042},  \pg{A30}.

\bibitem[Shimizu {\em et~al.\/}(1999)Shimizu, Minekawa, Hirose \&
  Kawase]{shimizu1999drop}
{\sc \au{Shimizu, K}, \au{Minekawa, K}, \au{Hirose, T} \& \au{Kawase, Y}}
  \yr{1999}  \at{Drop breakage in stirred tanks with newtonian and
  non-newtonian fluid systems}.  \jt{Chemical Engineering Journal}
  \bvol{72}~(2),  \pg{117--124}.

\bibitem[Solsvik \& Jakobsen(2015)]{solsvik2015single}
{\sc \au{Solsvik, Jannike} \& \au{Jakobsen, Hugo~A}} \yr{2015}  \at{Single drop
  breakup experiments in stirred liquid--liquid tank}.  \jt{Chemical
  Engineering Science}  \bvol{131},  \pg{219--234}.

\bibitem[Vah{\'e} \& Thiesset(2026)]{VaheThiesset2026}
{\sc \au{Vah{\'e}, J.} \& \au{Thiesset, Fabien}} \yr{2026}  \at{Conversions
  between kinetic and surface energy in periodically forced multiphase
  turbulence}.  \jt{Physical Review Fluids}  \bvol{11},  \pg{074002}.

\bibitem[Vankova {\em et~al.\/}(2007)Vankova, Tcholakova, Denkov, Ivanov,
  Vulchev \& Danner]{vankova2007}
{\sc \au{Vankova, Nina}, \au{Tcholakova, Slavka}, \au{Denkov, Nikolai~D},
  \au{Ivanov, Ivan~B}, \au{Vulchev, Vassil~D} \& \au{Danner, Thomas}} \yr{2007}
   \at{Emulsification in turbulent flow: 1. mean and maximum drop diameters in
  inertial and viscous regimes}.  \jt{J.\ Colloid Interface Sci.}
  \bvol{312}~(2),  \pg{363--380}.

\bibitem[Vashisth \& Andersson(2025)]{vashisth2025dynamics}
{\sc \au{Vashisth, Vikash} \& \au{Andersson, Ronnie}} \yr{2025}  \at{Dynamics
  of bubble breakup under turbulent flow conditions}.  \jt{Chemical Engineering
  Journal}  \bvol{511},  \pg{162019}.

\bibitem[Vela-Mart{\'{i}}n \& Avila(2021)]{AlbertoJFM}
{\sc \au{Vela-Mart{\'{i}}n, Alberto} \& \au{Avila, Marc}} \yr{2021}
  \at{Deformation of drops by outer eddies in turbulence}.  \jt{Journal of
  Fluid Mechanics}  \bvol{929},  \pg{A38}.

\bibitem[Vela-Mart{\'\i}n \& Avila(2022)]{AlbertoSciAdv}
{\sc \au{Vela-Mart{\'\i}n, Alberto} \& \au{Avila, Marc}} \yr{2022}
  \at{Memoryless drop breakup in turbulence}.  \jt{Science Advances}
  \bvol{8}~(50),  \pg{eabp9561}.

\bibitem[Vela-Mart{\'\i}n \& Avila(2024)]{vela2024large}
{\sc \au{Vela-Mart{\'\i}n, Alberto} \& \au{Avila, Marc}} \yr{2024}
  \at{Large-scale patterns set the predictability limit of extreme events in
  {Kolmogorov} flow}.  \jt{Journal of Fluid Mechanics}  \bvol{986},  \pg{A2}.

\bibitem[Wu {\em et~al.\/}(2026)Wu, Li \& Coletti]{Wu2026}
{\sc \au{Wu, Guotao}, \au{Li, Yaxing} \& \au{Coletti, Filippo}} \yr{2026}
  \at{Localised inter-scale energy cascades in free-surface turbulence}.
  \jt{{Journal of Fluid Mechanics}}  \bvol{1026},  \pg{A51}.

\bibitem[Xiao {\em et~al.\/}(2021)Xiao, Brillo, Lee, Hyers \&
  Matson]{xiao2021impact}
{\sc \au{Xiao, Xiao}, \au{Brillo, J{\"u}rgen}, \au{Lee, Jonghyun}, \au{Hyers,
  Robert~W} \& \au{Matson, Douglas~M}} \yr{2021}  \at{Impact of convection on
  the damping of an oscillating droplet during viscosity measurement using the
  iss-eml facility}.  \jt{npj Microgravity}  \bvol{7}~(1),  \pg{36}.

\bibitem[Yi {\em et~al.\/}(2021)Yi, Toschi \& Sun]{yi2021global}
{\sc \au{Yi, Lei}, \au{Toschi, Federico} \& \au{Sun, Chao}} \yr{2021}
  \at{Global and local statistics in turbulent emulsions}.  \jt{Journal of
  Fluid Mechanics}  \bvol{912},  \pg{A13}.

\bibitem[Yue {\em et~al.\/}(2007)Yue, Zhou \& Feng]{yue2007spontaneous}
{\sc \au{Yue, Pengtao}, \au{Zhou, Chunfeng} \& \au{Feng, James~J.}} \yr{2007}
  \at{Spontaneous shrinkage of drops and mass conservation in phase-field
  simulations}.  \jt{Journal of Computational Physics}  \bvol{223}~(1),
  \pg{1--9}.

\end{thebibliography}

\appendix

\section{Equilibrium surface energy}\label{ap:Hinf}
\begin{figure}
    \centering
    \includegraphics[width=\textwidth]{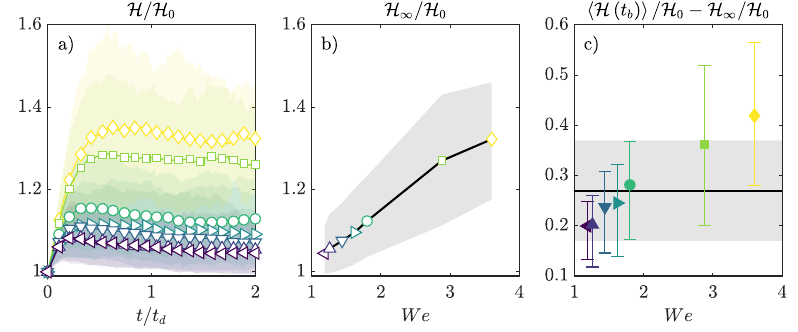}
    \caption{Equilibrium surface energy at different Weber numbers. Colours and symbols correspond to different Weber numbers (as in figure~\ref{fig:fig2}): blue low Weber number, yellow high Weber number. a) Ensemble-averaged surface energy at different time steps. b) Equilibrium surface energy at different Weber numbers. The shaded region covers the $10^{th}$ to $90^{th}$ percentile interval. c) Difference between the ensemble-averaged energy at breakup and the equilibrium surface energy. The horizontal black line indicates the value of $\xi_{c}=0.27$, and the shaded gray region covers a positive and negative standard deviation on the determination of $\xi_{c}$.}
    \label{fig:figHinit}
\end{figure}

%In \S\ref{sec:strain}, we model the energy transfer from interfacial to kinetic energy as a linear damping in the equation for $\mathcal{H}$. The damping is proportional to the difference between the surface energy and the equilibrium surface energy $\mathcal{H}_{\infty}$. In this appendix, we justify this modelling choice. 

In figure~\ref{fig:figHinit}a, we show the temporal evolution of the ensemble-averaged surface energy for the simulations carried out in this paper (table~\ref{tab:Database}). The ensemble average was computed as follows. For each Weber number and time step, we averaged the surface energy of all the drops that had not broken yet and were not going to break in the next eddy-turnover time. The ensemble-averaged surface energy reaches a statistically steady state within $t \lesssim t_d$. We computed the equilibrium surface energy, $\mathcal{H}_{\infty}$, as the ensemble- and time-averaged surface energy $\mathcal{H}$ during the time span $1.5~t_{d} \leq t \leq 2 ~t_{d}$. As shown in figure~\ref{fig:figHinit}b, $\mathcal{H}_{\infty}$ approaches $\mathcal{H}_0$ as $\mathrm{We}\rightarrow0$ and grows as $\mathrm{We}$ increases. In figure~\ref{fig:figHinit}c, the difference between the average breakup energy $\mathcal{H}(t_b)/\mathcal{H}_{0}$ and the average equilibrium energy $\mathcal{H}_{\infty}/\mathcal{H}_{0}$ is shown to remain close to $0.27$ at the low $\mathrm{We}<2$ of interest. Thus, our model captures the fact that breakup occurs (on average) when the difference between its surface energy and the equilibrium surface energy, normalised with $\mathcal{H}_0$, exceeds the threshold $\xi_c \approx 0.27$. 

\section{Outer stretching before breakup}\label{ap:Conditional}
\begin{figure}
    \centering
    \includegraphics[width=\textwidth]{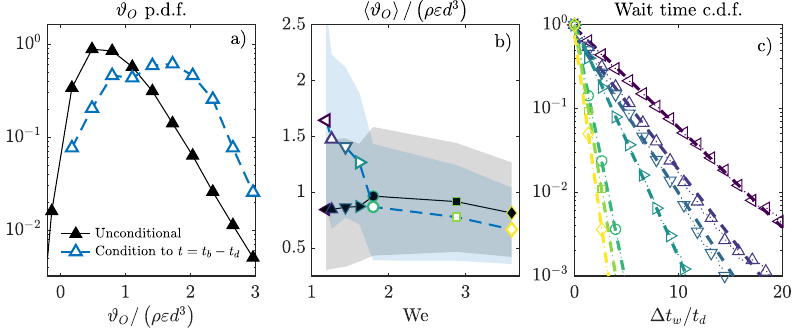}
    \caption{a) P.d.f.\ of the outer strain at $\mathrm{We}=1.3$ (solid line and black markers) and the p.d.f.\ of $\vartheta_{O}$ conditioned on one eddy-turnover time before breakup, $t=t_{b}-t_{d}$ (dashed blue line and white markers). b) Ensemble average of $\vartheta_{O}$ as a function of the Weber number. The solid black line and black markers show the unconditional ensemble average $\left \langle \vartheta_{O} \right \rangle$, whereas the dashed blue line and white markers show the average conditioned on $t=t_{b}-t_{d}$. The shaded region covers the $10^{th}$ to $90^{th}$ percentile interval. c) C.d.f.\ of the waiting times between outer-stretching events equal to or higher than the ensemble average $\vartheta_{O} \geq \left \langle \vartheta_{O} \right \rangle$ conditioned on $t=t_{b}-t_{d}$. The colours and symbols correspond to the same DNS carried out in the paper: from low (blue) to high (yellow).}
    \label{fig:figVO}
\end{figure}

In~\S\ref{sec:extreme}, we assumed that breakup is initiated by a rare, intense outer-stretching event of duration $t_{\vartheta_{O}}\approx 0.65t_d$. This is in qualitative agreement with \citet{cannon2026strain}, who showed that for $\mathrm{We}=1.82$ breakup is
preceded by high-intensity extensional strain events occurring approximately $t_d$ before fragmentation. In figure~\ref{fig:figVO}a, we show the conditional p.d.f.\ of the outer strain conditioned on one eddy-turnover time before breakup, $t=t_{b}-t_{d}$, for $\mathrm{We}=1.3$.  At this time, the p.d.f.\ is clearly biased toward strongly positive values (in comparison to the unconditional p.d.f.\ of $\vartheta_O$). We repeated this conditional analysis for all the Weber numbers in our database (table~\ref{tab:Database}). In figure~\ref{fig:figVO}b, we show that the ensemble-averaged outer strain at $t=t_{b}-t_{d}$ increases as the Weber number decreases. This further supports that, at low Weber numbers, breakup is triggered by intense outer-stretching events, while at higher Weber numbers, the outer strain does not need to be higher than average to trigger breakup. 
In figure~\ref{fig:figVO}c, we show the c.d.f.\ of waiting times of intense outer-stretching events with intensity higher than the ensemble average $\left \langle \vartheta_{O} \right \rangle$ conditioned on $t=t_{b}-t_{d}$ (the data points in figure~\ref{fig:figVO}b). The p.d.f.s are exponential, consistent with a Poisson process. Their decay rate increases with the Weber number, i.e. as the Weber number decreases, the mean waiting time increases and the highly intense events preceding the breakup become rarer. We repeated the analysis of figure~\ref{fig:figVO}, but conditioning on $t=t_{b}-t_{\sigma}$ and $t=t_{b}-0.5~t_{d}$. The results are similar and not shown here.

\section{Comparison with the model by \citet{coulaloglou1977description}}\label{ap:CT77}

\begin{figure}
    \centering
    \includegraphics[width=\textwidth]{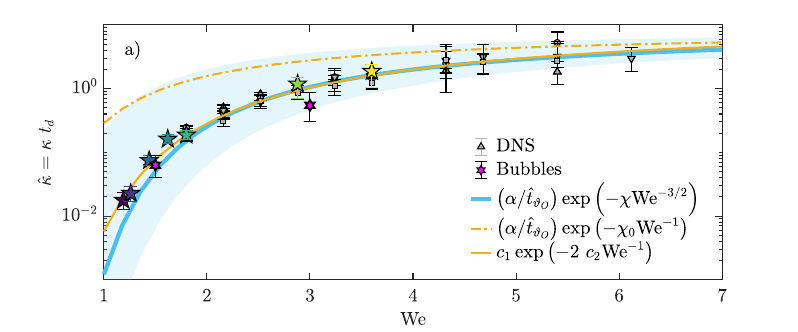}
    \caption{As figure~\ref{fig:fig5}b but including curves for the model by \citet{coulaloglou1977description}, equation~\eqref{eq:C77}. The thin yellow solid line corresponds to the two-parameter least-squares fit by \citet{AlbertoSciAdv}, with $c_{1}\approx14.8$ and $c_{2}\approx3.9$. The thin dashed-dotted line corresponds to the same formula but with $c_2=\chi_{0}/2$, $\chi_{0}=\gamma~\pi~\xi_{c}/\hat{t}_{\vartheta_{O}}$, and $c_1=\alpha/\hat{t}_{\vartheta_{O}}$, i.e. the limit of our model for negligible damping, equation~\eqref{eq:br_infy}.}
    \label{fig:figAp2}
\end{figure}

The model of \citet{coulaloglou1977description}, equation~\eqref{eq:C77}, rests upon two key underlying hypotheses regarding the time scales of the relevant physical processes. In particular, it assumes that the relaxation time is (i) longer than the time scale of energy injection by the energetic eddies, and (ii)  shorter than the time between encounters with sufficiently strong eddies. The second hypothesis is clearly satisfied at low Weber numbers (\S\ref{sec:lowWe_sol}). However, the first hypothesis is not: the drop relaxes at the capillary time scale, which at low Weber numbers is shorter than the duration of the stretching by outer eddies, which acts at the eddy-turnover time scale. In fact, both hypotheses cannot hold simultaneously, because the time scale of energy injection and that of encounters with sufficiently strong eddies are the same, $\hat{t}_{\vartheta_{O}}\approx0.65$. Still, the excellent agreement of a fit of equation~\eqref{eq:C77} (shown in figure~\ref{fig:figAp2} as a thin solid line) to the numerical breakup data warrants a detailed comparison of both models. First, we note that equation~\eqref{eq:C77} has two fit parameters, a multiplicative constant, $c_1$, and a constant in the exponent, $c_2$, that must be fitted to breakup data, whereas the parameters of our model are inferred from DNS without fitting them to breakup rates. Second, in our model, the analogue of $c_2$, i.e. $\chi=\gamma\,\pi \, \xi_c  \,c_d/(1-\delta)$  (equation~\eqref{eq:chi}), depends itself on the Weber number through $\delta=\exp(-c_d \, \hat{t}_{\vartheta_{O}}\,\mathrm{We}^{-1/2})$ (equation~\eqref{eq:delta}). 

Let us consider the limit $c_d \hat{t}_{\vartheta_{O}}/\sqrt{\mathrm{We}} \rightarrow  0$ in our model, which precisely corresponds to the assumption of the damping being much slower than energy injection \citep{coulaloglou1977description}. In this limit, $1-\delta\approx c_d \hat{t}_{\vartheta_{O}}/\sqrt{\mathrm{We}}$, and our model naturally recovers theirs,
\begin{equation}\label{eq:br_infy}
\hat{\kappa}_{c_{d} \rightarrow 0} \approx c_1\, \exp\left(-2 c_2 \, \mathrm{We}^{-1}\right),
\end{equation}
with $c_2=\chi_{0}/2=\gamma~\pi~\xi_{c}/\hat{t}_{\vartheta_{O}}/2\approx 1.7$ and $c_1=\alpha/\hat{t}_{\vartheta_{O}}$. This is shown as a dashed-dotted line in figure~\ref{fig:figAp2} and approaches the full model breakup prediction as $\mathrm{We}\rightarrow \infty$ and damping becomes negligible. In fact, the limit $\mathrm{We}\rightarrow \infty$ is identical to the limit $c_d \hat{t}_{\vartheta_{O}}/\sqrt{\mathrm{We}} \rightarrow  0$. As $\mathrm{We}$ decreases, the damping rapidly gains importance and the two curves clearly diverge. Only by treating the constants $c_1$ and $c_2$ as independently fitted parameters, as done in the experimental and numerical literature, can the breakup data be matched (solid thin curve in figure~\ref{fig:figAp2}).

\end{document}